%% file: kcmu.tex
\documentclass[journal]{IEEEtran}

\usepackage{cite}
\usepackage{amsmath,amssymb,amsfonts}
\usepackage[T1]{fontenc}
\usepackage{graphicx}
\graphicspath{{figures/mpl/}{figures/}{tables/}}
\usepackage{booktabs}
\usepackage{multirow}
\usepackage{tabularx}
\usepackage{threeparttable}
\usepackage{stfloats}
\usepackage{algorithm}
\usepackage{algorithmic}
\usepackage{subcaption}
\usepackage[table]{xcolor}
\usepackage{hyperref}
\usepackage{xspace}

\hypersetup{colorlinks=true,linkcolor=blue,citecolor=blue,urlcolor=blue}

\newcommand{\unison}{\textsc{Unison}\xspace}
\newcommand{\spear}{\textsc{Spear}\xspace}
\newcommand{\tide}{\textsc{Tide}\xspace}
\newcommand{\kcmu}{\unison}
\newcommand{\esp}{\spear}
\newcommand{\okt}{\tide}

\title{UNISON: A Co-Designed Near-Memory Scheduler of Session KV Residency for LLM Agents}

\author{Fan~He,
        Yan~Li,~\IEEEmembership{Member,~IEEE,}
        and~Xiaoyang~Zeng,~\IEEEmembership{Senior~Member,~IEEE}%
\thanks{This work was supported by the National Natural Science Foundation of China under Grant 62574049. (\textit{Corresponding author: Yan Li.})}%
\thanks{Fan He, Yan Li, and Xiaoyang Zeng are with the State Key Laboratory of Integrated Chips and Systems, Fudan University, Shanghai 200433, China (e-mail: liyan@fudan.edu.cn).}%
}

\begin{document}

\maketitle

\input{sec_abstract}

\IEEEpeerreviewmaketitle

\input{sec_intro}
\input{sec_related}
\input{sec_method}
\input{sec_arch}
\input{sec_sweval}
\input{sec_hweval}
\input{sec_conc}

\bibliographystyle{IEEEtran}
\bibliography{refs}

\end{document}

%% file: sec_abstract.tex
\begin{abstract}
Large language models are increasingly composed into agent loops that
plan, call tools, and resume the same task after each action.
These loops press a shared memory hierarchy harder than conventional
multi-turn chat, because they hold a growing key-value (KV) prefix
across tool waits and place many sessions on one static
random-access memory (SRAM) and high-bandwidth memory (HBM) pool,
so that eviction and hierarchical placement become a session-level
efficiency problem orthogonal to compute-mode optimization.
Existing proxies based on recency, timeout, or identity miss the
mechanism information of the loop and therefore treat a live wait
as a cold, discardable unit. To
address this problem, we present Unified Native Inter-turn Session
Orchestration Nexus (\unison), an event-driven near-memory scheduler beside the
memory hierarchy in which Survival-Penalty Eviction for Agent
Return-gap (\spear) and Tiering in Idle-window DMA Events (\tide)
share one live ranking. \spear selects who leaves from a gap average
and a turn-indexed hazard, while \tide spends the observed wait as a
direct memory access (DMA) budget for who sits in the fast tier. On
both coding and general-mission benchmarks with three model
families, totaling 1\,415 sessions and 33\,596 turns, the joint policy
is the best non-oracle entry on hit rate and average memory access
time (AMAT) on every trace, raising the hit rate by $0.3\%$ to
$23.1\%$ and reducing AMAT by $22\%$ to $51\%$, and lowering time to
first token (TTFT) by $58\%$ to $89\%$ on the long-horizon traces. A structural necessity analysis shows that the unified near-memory
design cannot be decomposed into independent IPs or realized in
software without re-introducing documented failure modes. The 28-nm
CMOS scheduling core occupies 0.169\,mm$^{2}$ at 13.6\,mW and
150\,MHz, a negligible overhead relative to the KV
hierarchy it manages, and reproduces the floating-point ranking at a
Kendall $\tau$ exceeding 0.998.
\end{abstract}


%% file: sec_intro.tex
\section{Introduction}
\label{sec:intro}

\IEEEPARstart{L}{arge} language models (LLMs) are now widely deployed
as general-purpose generators, and the rise of tool use has composed
them into multi-step agent
loops~\cite{yao_react_2023,wang_survey_2024} that plan, invoke
external tools, and resume the same task after each action. Although
multi-turn chat already retains a conversation prefix in the
key-value (KV) cache, an agent loop further superimposes structured
tool waits, larger tool-result inserts, and more model invocations
per task, so that the object of serving is no longer a single request
but a session that is expected to return.

\begin{figure}[t]
\centering
\includegraphics[width=\columnwidth]{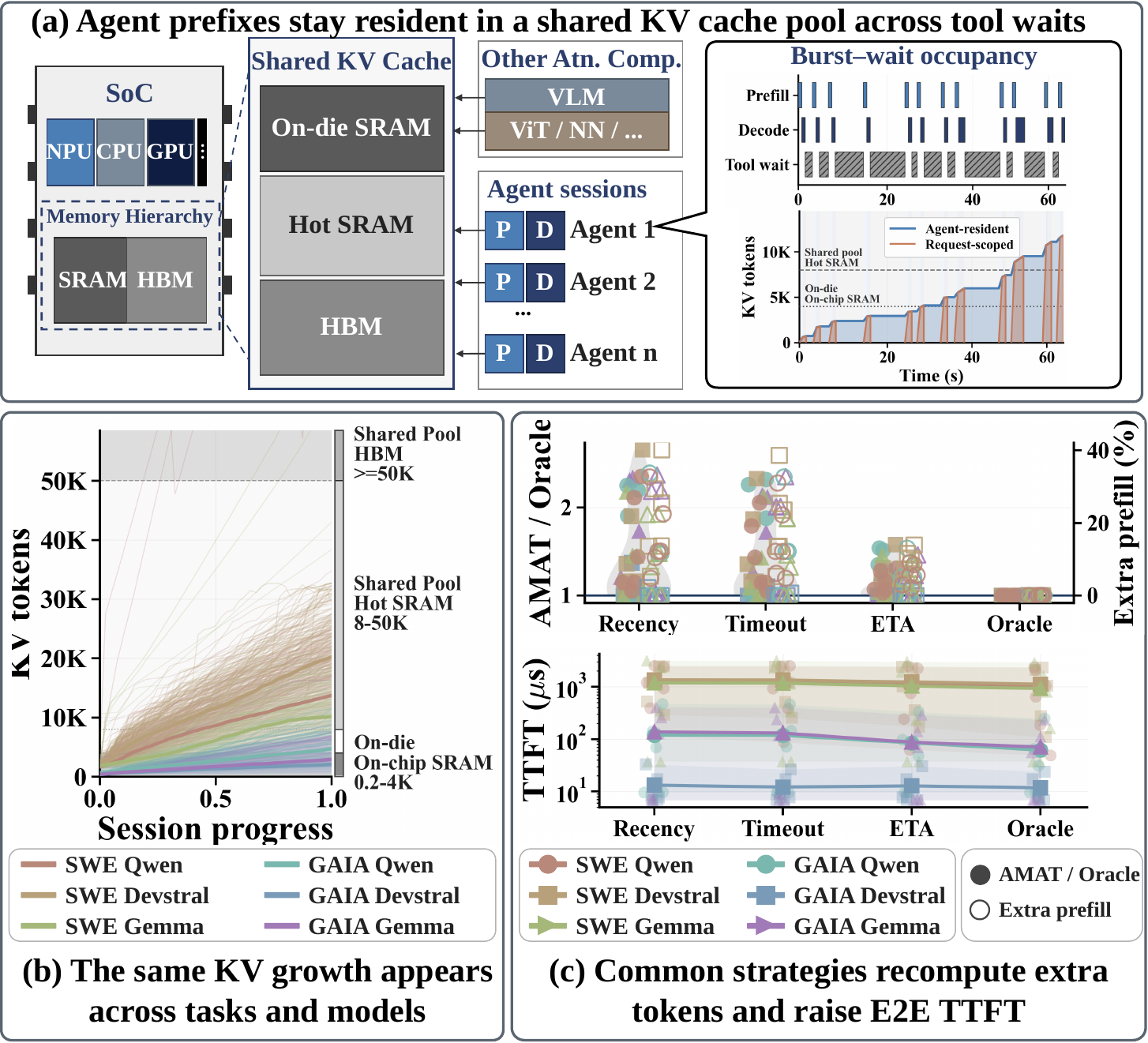}
\caption{Motivation. \textbf{(a)}~Agent-resident occupancy versus a
request-scoped counterfactual on the same twelve-turn session.
\textbf{(b)}~Monotonic prefix growth across 1\,415 sessions into edge
and cloud SRAM/HBM envelopes. \textbf{(c)}~Recency, timeout, and ETA
remain above a B\'el\'ady oracle in AMAT, extra prefill, and TTFT.}
\label{fig:motivation}
\end{figure}

Serving systems have accordingly devoted most of their effort to how
tokens are computed, including how prefill and decode share or split
devices~\cite{patel_splitwise_2024,zhong_distserve_2024}. Under agent
loops, however, the end-to-end cost is often set by how the KV cache
is retained, evicted, and placed. Whereas a multi-turn conversation
remains paced by one user, agent workloads place many agents and many
sessions on the same pool at once and hold a growing prefix across
tool waits. The resulting live set shares static random-access memory (SRAM) and
high-bandwidth memory (HBM) with other attention clients, including
prefill, decode, and an optional vision-language model, on both edge
systems-on-chip (SoCs) and cloud servers. Once aggregate residency
outgrows that pool, eviction and hierarchical placement become an
efficiency problem in their own right and stand orthogonal to
compute-mode optimization. Fig.~\ref{fig:motivation}(a) records this
operating point.

Fig.~\ref{fig:motivation}(a) contrasts two KV contracts on the same
twelve-turn session. The agent-resident curve holds the prefix through
every tool wait, whereas the request-scoped pulses drop the cache
after each decode. Unlike a multi-turn chat staircase whose idle
interval is a human pause that may never return, an agent tool wait is
a scheduled return, and recency or timeout policies therefore treat the
second kind of idle as if it were the first.

That resident contract becomes expensive once the prefix outgrows the
finite pool. Fig.~\ref{fig:motivation}(b) overlays prefix tokens on
coding and web-agent traces across 1\,415 sessions and three model
families. The curves rise monotonically into the SRAM and HBM
envelopes that edge and cloud stacks provision, and a budget that
cannot hold the live set is what makes session-level scheduling
unavoidable.

Published proxies miss the mechanism information that the loop
already produces and therefore treat a live wait as a cold,
discardable unit. Recency and timeout invert the rank by reading a
tool wait as coldness. A per-request lifecycle issues an absolute
keep-or-drop without ever ranking the live set against other
residents~\cite{abhyankar_infercept_2024,li_continuum_2026}. A
tool-type arrival table ranks residents yet assigns the same wait to
every job that shares a tool. Identity- or workflow-conditioned
scores separate residents only while roles or graphs remain diverse
and flatten toward recency once the batch is
uniform~\cite{cachescout_2026,pan_kvflow_2025,zheng_pbkv_2026}.
Hierarchical stores enlarge the pool yet still place from a static
plan, a job-scheduler hint, or prefix popularity, so that a kept
prefix is served on the slow path. Paged and radix
runtimes~\cite{kwon_pagedattention_2023,zheng_sglang_2024} manage the
pool at block granularity. Token, device, and interconnect engines
decide which bytes of one request to compute or move, a decision
that is complementary to ranking who stays in the shared pool, as
Table~\ref{tab:hw-kv} places on one grid. As Fig.~\ref{fig:motivation}(c)
reports, recency, timeout, and estimated time of arrival (ETA) all
sit above a B\'el\'ady next-reference oracle in average memory
access time (AMAT), extra prefill, and time to first token (TTFT)
across six traces and six capacity envelopes. Because that oracle is not observable online, a
useful substitute must exploit a signal that is already present on
every session, observable at runtime, and cheap to maintain, namely
the session's own gap history and its own progress toward completion.

Orthogonal to token datapaths and cluster routers that pursue the
same end-to-end efficiency from complementary vantage points, this
work proposes a near-memory scheduler beside the hierarchy
for long-horizon agent tasks in which many attention clients compete
for a two-tier pool on the edge and in the cloud. The scheduler
must decide who leaves the pool and who sits in the fast tier, and
both decisions are conditioned on the loop's mechanism information
rather than on an attempt to read task content.

To address those issues, we present
\textbf{U}nified \textbf{N}ative \textbf{I}nter-turn
\textbf{S}ession \textbf{O}rchestration \textbf{N}exus (\unison),
an event-driven near-memory scheduler in which
\textbf{S}urvival-\textbf{P}enalty \textbf{E}viction for
\textbf{A}gent \textbf{R}eturn-gap (\spear) and
\textbf{T}iering in \textbf{I}dle-window \textbf{D}MA
\textbf{E}vents (\tide) share one live ranking in a session
register file beside the memory hierarchy.
\spear scores remaining distance from a gap exponential moving
average (EMA) and a turn-indexed hazard and selects who leaves, while
\tide reads the complementary order, spends the observed wait as a
direct memory access (DMA) budget, and selects who sits in the appropriate memory hierarchy.
The main contributions are as follows.
\begin{enumerate}
\item \textbf{Algorithm.} \spear and \tide jointly rank sessions from
loop signals that the serving runtime already observes. Evaluated on
six traces that cross the
SWE-bench~\cite{jimenez_swebench_2024} and
GAIA~\cite{mialon_gaia_2023} agent benchmarks with three model
families~\cite{qwen3_2025,devstral_2025,gemma4_2026}, totaling 1\,415
sessions, 33\,596 turns, and six capacity envelopes from edge to
cloud, the joint policy is the best non-oracle entry on hit rate and
AMAT on every trace, raising the hit rate by $0.3\%$ to $23.1\%$ and reducing AMAT by
$22\%$ to $51\%$. On the long-horizon
traces it also lowers serving TTFT by 58\% to 89\% when extra
prefill sits on the critical path.
\item \textbf{Architecture.} We show that cycle-level event fidelity,
unified eviction and migration state, and deterministic gap-window
completion jointly require a near-memory control plane that cannot be
decomposed into independent IPs or realized in software. The 28-nm
CMOS scheduling core occupies 0.169\,mm$^{2}$ at 13.6\,mW and 150\,MHz for 64
sessions, a negligible overhead that reproduces the floating-point
ranking at Kendall $\tau{>}0.998$.
\end{enumerate}

%% file: sec_related.tex
\section{Related Work}
\label{sec:related}

A finite two-tier pool poses two ordering problems: eviction ranks who
leaves, and placement ranks who sits in the fast tier. Both oracles
reduce to B\'el\'ady next-reference distance, which is not observable
online, so every policy substitutes a proxy conditioned on progressively
richer information.

\subsection{Information-Conditioned Eviction Methods}
\label{sec:related-evict}

Access-history methods treat a recent touch as evidence of future need.
PagedAttention~\cite{kwon_pagedattention_2023} pages a growing KV pool
in the style of virtual memory, and
SGLang~\cite{zheng_sglang_2024} reuses shared prefixes through a radix
tree. Both evict by block recency, so a session alive on a tool wait
appears oldest and leaves first.

Lifecycle methods promote the pause to a first-class request state.
InferCept~\cite{abhyankar_infercept_2024} chooses discard, preserve, or
swap from a per-request waste account, and
Continuum~\cite{li_continuum_2026} pins paused KV with a time-to-live
derived from reload cost. Both keep a paused session resident yet issue
an absolute keep-or-drop for one request, so the locally cheap choice
can be the globally expensive victim.

Arrival-table methods predict the next request.
AGSERVE~\cite{agserve_2025} reads the next arrival from a tool-type
latency table, assigning the same wait to every session that shares a
tool, which homogenizes heterogeneous jobs and mis-ranks the longest
residents.

Structure-conditioned methods exploit workflow or role diversity.
CacheScout~\cite{cachescout_2026} learns a Markov table over agent
roles, KVFlow~\cite{pan_kvflow_2025} reads steps-to-execution off a
declared graph, and PBKV~\cite{zheng_pbkv_2026} predicts future agent
calls. The contrast in all three is borrowed from structural diversity;
a homogeneous batch flattens the scores and the policy falls back to
recency.

A parallel line shortens the token set of one trajectory rather than
ranking who leaves a shared pool.
IntentKV~\cite{li_intentkv_2026} scores history tokens from cross-turn
intent, MemDecay~\cite{matam_memdecay_2026} assigns region-specific
decay, and CommitKV~\cite{huang_commitkv_2026} retires pages only after
a tool-call commit. All three trade accuracy for a smaller working set
without comparing live sessions against one another.
Semantic queues~\cite{fang_saecache_2026}, variable-grain
paging~\cite{jeon_granikv_2026}, and workflow-aware job
admission~\cite{ni_topas_2026} similarly reshape what a request
contains or which job runs, yet none ranks live sessions on a shared
pool.

Learned replacement policies trained by reinforcement learning or
imitation, such as LRB~\cite{song_lrb_2020} in web caching, achieve
strong offline accuracy but require per-decision inference times of
hundreds of microseconds to milliseconds, which exceeds the
single-digit microsecond budget of a near-memory scheduler by two
orders of magnitude and therefore cannot serve as an online eviction
oracle in the hardware pipeline targeted here.

Taken together, access history inverts the ranking, a per-request
lifecycle yields no ranking, a population table homogenizes it, and
borrowed structure yields a ranking that vanishes when the batch is
uniform, and none of these proxies conditions on the signal that every
session carries in every regime, namely its own observed gaps and its own
progress toward completion.

\subsection{Hierarchy-Aware Placement Strategies}
\label{sec:related-place}

A kept session still has an access cost; the oracle places the soonest
next reference in the fast tier. Prior placement work has widened its
information from a static offload plan to job-scheduler hints to a
global reuse store.
FlexGen~\cite{sheng_flexgen_2023} solves a linear program over GPU,
CPU, and disk, conditioning on membership in the current computation, so
a session alive on a tool wait is demoted first.
CachedAttention~\cite{gao_cachedattention_2024} places blocks from
scheduler hints across a three-level store, but the hint names the next
scheduled job, not the remaining wait of each session.
Mooncake~\cite{qin_mooncake_2025} pools DRAM, SSD, and RDMA under
prefix-popularity ordering, which ranks by reuse frequency rather than
by proximity.

Consequently, a static plan inverts the placement in time, a scheduler
hint homogenizes it across pauses, and a popularity score orders it by
frequency instead of proximity, so all three keep the right bytes in
the wrong place because none conditions on the gap this session just
opened.

\subsection{Near-Memory and Datapath KV Hardware}
\label{sec:related-hw}

Token, device, and interconnect engines decide which bytes of one
request to compute or move, a decision complementary to ranking who
stays in a shared pool. Table~\ref{tab:hw-kv} in
Section~\ref{sec:hweval} places both objects on one grid.

Importance datapaths prune or skip tokens inside one decode step.
Token-Picker~\cite{park_tokenpicker_2024} withholds transfers based on
pre-softmax probability, UniCAIM~\cite{xu_unicaim_2025} prunes via
content-addressable search, a KV-MMU~\cite{moradifirouzabadi_kvmmu_2025}
replaces least-relevant tokens at runtime,
HiKV~\cite{fang_hikv_2026} evicts tokens then loads significant
elements through a reconfigurable sorter, and
Kelle~\cite{xia_kelle_2025} evicts by attention score in embedded DRAM.
Look-back and voting designs such as MATA~\cite{zhu_mata_2026} and
VEDA~\cite{wang_veda_2025} belong to the same datapath family.
Capacity-expansion hardware enlarges the store.
CXL-SpecKV~\cite{liu_cxlspeckv_2026} offloads KV to FPGA memory
with speculative prefetch, and V-Rex~\cite{kim_vrex_2026} retrieves
clustered video-token subsets. None of these engines observes a tool
wait or a pool of other live sessions.

A near-memory scheduler sits beside the hierarchy and ranks session
residency on request and gap events, so that eviction and tiering
become two reads of that ranking. The datapath can still prune tokens
of the request now in flight; the two mechanisms do not substitute for
each other.

%% file: sec_method.tex
\section{Methodology}
\label{sec:method}

\begin{figure*}[t]
\centering
\includegraphics[width=\textwidth]{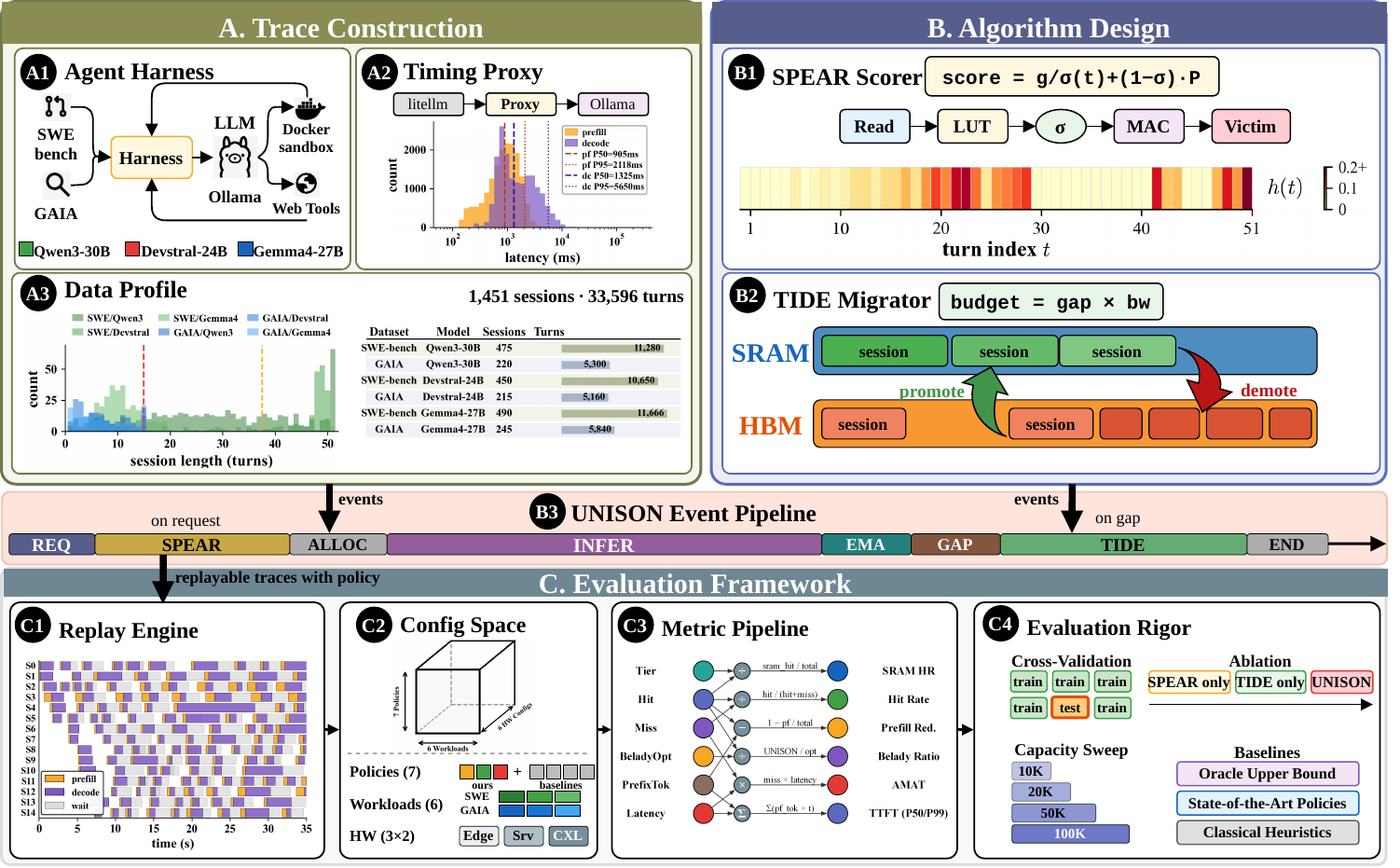}
\caption{\unison methodology. A shared event log bridges workload
characterization, survival-based scoring, idle-window migration, and
multi-metric evaluation.}
\label{fig:method}
\end{figure*}

This section decomposes the online control problem of
Section~\ref{sec:intro} into three sub-problems and solves each under
causal information alone. Section~\ref{sec:problem} formalizes the
two-tier pool and the cost asymmetry that makes victim choice the
dominant lever. Section~\ref{sec:esp} introduces \spear, which
approximates the B\'el\'ady eviction oracle from two observable
signals, namely a gap recurrence estimate and a turn-indexed
completion hazard. Section~\ref{sec:okt} introduces \tide, which converts
measured idle windows into tier-placement migrations ranked by the
same signal. Section~\ref{sec:loop} unifies both mechanisms in a
single event-driven control loop as shown in Algorithm~\ref{alg:unison}.
Fig.~\ref{fig:method} gives the corresponding pipeline.

\subsection{Workload Model and Problem Formulation}
\label{sec:problem}

An agent execution is a session $s$ of $T_s$ turns. Turn $k$ arrives at
wall time $a_{s,k}$ after a tool gap
\begin{equation}
  g_{s,k}=a_{s,k}-c_{s,k-1},
  \label{eq:gap}
\end{equation}
where $c_{s,k-1}$ is the completion time of the previous turn.
The resident KV length $\ell_{s,k}$ is prefix-heavy because later turns
inherit most of their context from earlier ones, so a hit avoids
re-prefilling the full prefix while a miss pays the entire sequence
length. The reuse ratio
\begin{equation}
  \rho_{s,k}
  =\frac{\ell_{s,k-1}}{\ell_{s,k}},
  \qquad
  \Delta\ell_{s,k}
  =\ell_{s,k}-\ell_{s,k-1}
  \label{eq:reuse}
\end{equation}
quantifies this asymmetry. A miss costs
$\ell_{s,k}\,t_{\mathrm{p}}$ whereas a hit costs only
$\Delta\ell_{s,k}\,t_{\mathrm{p}}$, making victim choice the
dominant lever on serving latency.

The live set $\mathcal{L}(t)$ occupies a two-tier budget of SRAM
capacity $C_{\mathrm{S}}$ and HBM capacity $C_{\mathrm{H}}$ together
with a hard session-slot limit $N_{\mathrm{sess}}$,
\begin{equation}
  \sum_{s\in\mathcal{L}_{\mathrm{S}}}\ell_s\le C_{\mathrm{S}},
  \qquad
  \sum_{s\in\mathcal{L}_{\mathrm{H}}}\ell_s\le C_{\mathrm{H}},
  \qquad
  \lvert\mathcal{L}_{\mathrm{S}}\cup\mathcal{L}_{\mathrm{H}}\rvert
    \le N_{\mathrm{sess}}.
  \label{eq:cap}
\end{equation}
A policy maps $\mathcal{L}(t)$ to a victim on overflow and to
promotion and demotion candidates during a gap. The offline B\'el\'ady
oracle $v^{\star}$ evicts the session whose next request is farthest
in the future. An online policy $\pi$ may use only the causal
filtration $\mathcal{F}_{t}$, comprising the last gap, the turn index,
and the completion flag, and may use neither future arrivals nor agent-role
identity. The design target is to approximate $v^{\star}$ while
remaining a function of $\mathcal{F}_{t}$ alone. A learned policy
such as a small neural network could in principle fit the same
mapping, but the eviction decision sits on the critical path of every
cache miss and must complete within single-digit microseconds at the
hardware event rate, a latency regime that is incompatible with even a
minimal inference pass and that motivates the closed-form scoring
function developed below.

\subsection{\spear: Survival-Penalty Eviction for Agent Return-Gap}
\label{sec:esp}

The B\'el\'ady distance of a paused session depends on two quantities
that operate on different timescales, namely how long the current
tool gap will last and whether the session will return at all. \esp
approximates the first through a gap exponential moving average and
the second through a turn-indexed hazard, combining them into a
single eviction score that ranks all residents by estimated
next-reference distance.

\paragraph{Gap recurrence estimate}
The gap EMA tracks the characteristic return interval of session~$s$
with coefficient $\alpha{=}3/10$ as
\begin{equation}
  g_s
  \leftarrow
  \alpha\,g_{s,k}+(1-\alpha)\,g_s
  =\frac{3\,g_{s,k}+7\,g_s}{10}.
  \label{eq:ema}
\end{equation}
A session with a long smoothed gap is expected to remain idle longer,
so it should be evicted before a session that returns frequently.

\paragraph{Turn-indexed completion hazard}
A session closer to its final turn is more likely to complete and
release its KV permanently. The discrete hazard at turn $t$ is
\begin{equation}
  h(t)=\frac{d(t)}{n(t)},
  \qquad
  t\in\{0,\ldots,50\},
  \label{eq:hazard}
\end{equation}
where $n(t)$ sessions remain live at turn $t$ and $d(t)$ complete
there, counted over the observed population. The survival mass is the
local complement
\begin{equation}
  \sigma(t)
  =\max\bigl(\varepsilon,\,1-h(\min(t,50))\bigr),
  \qquad
  \varepsilon=0.01,
  \label{eq:surv}
\end{equation}
which requires a single table lookup, avoiding the sequential
dependence of the cumulative product-limit estimator.

\paragraph{Composite eviction score}
The two signals combine into
\begin{equation}
  \mathrm{score}(s)
  =
  \begin{cases}
    S_{\max},
      & s\text{ completed},\\[2pt]
    \dfrac{g_s}{\sigma(t)}+\bigl(1-\sigma(t)\bigr)P,
      & \text{otherwise},
  \end{cases}
  \label{eq:esp}
\end{equation}
where $P$ is a programmable completion penalty. A high score denotes
high eviction priority. The first term is the expected remaining wait
scaled by survival likelihood, so a session with low $\sigma$
near completion sees an inflated ratio that raises its eviction
priority. The
second term adds a penalty that grows as $\sigma$ falls, accelerating
eviction of sessions whose KV will soon have no future request. The
signal contribution and parameter sensitivity of~\eqref{eq:esp} are
evaluated in Sections~\ref{sec:sw-signal} and~\ref{sec:sw-main}.

Scoring determines who leaves the pool but not where a surviving
session should reside within the two-tier hierarchy. The
complementary placement problem is addressed next.

\subsection{\tide: Tiering in Idle-Window DMA Events}
\label{sec:okt}

A tool gap is wasted time from the perspective of the memory
hierarchy, because the paused session holds a tier slot but issues
no requests. \okt converts this idle window into a DMA budget and
migrates sessions between SRAM and HBM so that the lowest-scoring
residents occupy the fast tier when the next request arrives.

On a \texttt{gap\_start} event with estimated remaining duration
$\Delta$, the migration budget in tokens is
\begin{equation}
  \mathcal{B}=\Delta\cdot B,
  \label{eq:budget}
\end{equation}
where $B$ is the logical DMA bandwidth in tokens per nanosecond.
Let $\mathcal{H}$ be the HBM residents and $\mathcal{S}$ the SRAM
residents. Promotion and demotion then solve the complementary
selections
\begin{align}
  p
  &=\arg\min_{s\in\mathcal{H}}\mathrm{score}(s),
  \label{eq:promote}
  \\
  d
  &=\arg\max_{s\in\mathcal{S}}\mathrm{score}(s),
  \label{eq:demote}
\end{align}
subject to $\ell_p+\ell_d\le\mathcal{B}$ and to~\eqref{eq:cap} after
the swap. The key property of \okt is that it ranks migration
candidates with the same \esp score used for eviction, so eviction
quality directly determines placement quality, and the two mechanisms
share a single ranking signal rather than maintaining independent
state. This coupling is validated quantitatively in the component
ablation of Section~\ref{sec:sw-comp}.

\subsection{Joint Online Control Loop}
\label{sec:loop}

\esp and \okt share a single session register file updated on every
event. Algorithm~\ref{alg:unison} specifies the complete loop. Each
incoming event updates the register file, triggers an eviction scan
if capacity is violated, and initiates migrations if a gap is open
and the DMA engine is free. The two mechanisms read the same scores
from the same state, which is why a unified control plane
outperforms dual independent engines with periodic synchronization,
as the architectural ablation of Section~\ref{sec:arch}
demonstrates.

\begin{algorithm}[t]
\caption{\unison event loop}
\label{alg:unison}
\begin{algorithmic}[1]
\REQUIRE event $e$, live set $\mathcal{L}$, capacities
         $(C_{\mathrm{S}},C_{\mathrm{H}},N_{\mathrm{sess}})$,
         shared register file $R$, DMA state
\ENSURE  victim $v$ and logical DMA descriptors
\STATE write $e$ into $R$
         \COMMENT{turn, gap, completion}
\IF{$e$ is \texttt{gap\_end}}
  \STATE $g_s\leftarrow\alpha\,g_{s,k}+(1-\alpha)\,g_s$
         \COMMENT{\eqref{eq:ema}}
\ENDIF
\IF{$\mathcal{L}$ violates~\eqref{eq:cap}}
  \FOR{$s\in\mathcal{L}$}
    \STATE $\mathrm{score}(s)\leftarrow$~\eqref{eq:esp}
  \ENDFOR
  \STATE $v\leftarrow\arg\max_{s}\mathrm{score}(s)$
  \STATE evict $v$ from its tier
\ENDIF
\IF{$e$ is \texttt{gap\_start} and DMA is free}
  \STATE $\mathcal{B}\leftarrow\Delta\cdot B$
  \WHILE{$\mathcal{B}$ covers a legal pair}
    \STATE $p\leftarrow\arg\min_{s\in\mathcal{H}}\mathrm{score}(s)$
    \STATE $d\leftarrow\arg\max_{s\in\mathcal{S}}\mathrm{score}(s)$
    \STATE migrate $(p,d)$;
           $\mathcal{B}\leftarrow\mathcal{B}-\ell_p-\ell_d$
  \ENDWHILE
\ENDIF
\end{algorithmic}
\end{algorithm}

Section~\ref{sec:arch} maps this algorithmic contract to a
synthesizable fixed-point pipeline, and
Sections~\ref{sec:sweval}--\ref{sec:hweval} evaluate the resulting
policy against the B\'el\'ady oracle and competing baselines.

%% file: sec_arch.tex
\section{Architecture}
\label{sec:arch}

The policy of Section~\ref{sec:method} is defined over causal
information alone, but a software implementation on top of a
block-level cache cannot reconstruct the precise event timing that the
scoring contract requires, as the vLLM study in
Section~\ref{sec:sw-vllm} will confirm. A hardware realization
addresses three structural limitations that software cannot.
First, the controller must observe every request arrival, gap onset,
and gap closure with cycle-level fidelity, which a near-memory
placement provides without interrupt or polling overhead.
Second, eviction and migration decisions must complete within the
tool-gap window so that the promoted KV is already in the fast tier
when the next request arrives, which demands a deterministic pipeline
rather than a software thread.
Third, eviction and tiering are two reads of one ranking, and a single
register file eliminates the staleness that any dual-engine design
would introduce.

\begin{figure*}[t]
\centering
\includegraphics[width=\textwidth]{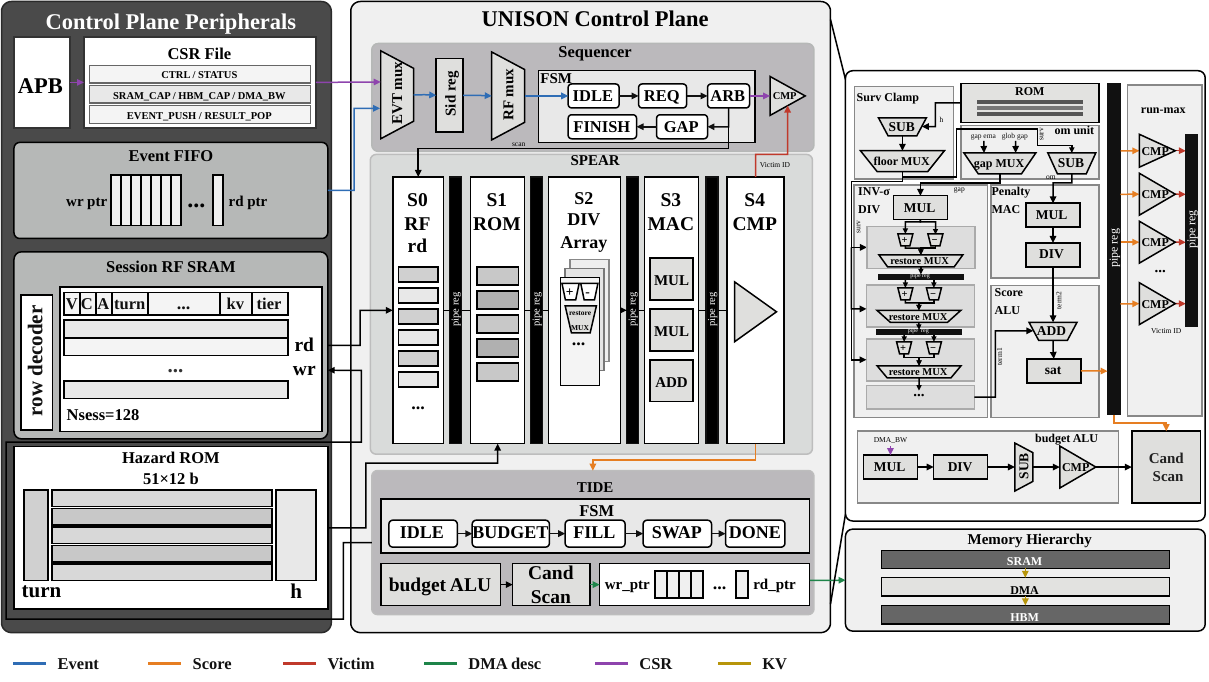}
\caption{\unison hardware architecture. The scheduling core integrates
\spear ranking and \tide migration into a unified pipeline beside the
two-tier KV hierarchy.}
\label{fig:arch}
\end{figure*}

This section maps the algorithmic policy to a synthesizable
control-plane IP. Section~\ref{sec:place} defines the controller
scope and system interface. Section~\ref{sec:org} derives the unified
control plane from the interleaving structure of agent workloads.
Section~\ref{sec:quant} locks the integer contract through a format
sweep. Section~\ref{sec:micro} maps the result to a pipelined event
core. Section~\ref{sec:dse} explores the session-capacity design
space and selects the design point. Fig.~\ref{fig:arch} shows the
resulting architecture.

\subsection{Controller Scope and System Placement}
\label{sec:place}

\kcmu is a session-level control-plane IP that ranks residency in a
shared KV pool. It does not compute attention, store KV payload, or
manage the HBM PHY. Runtime software writes events, namely request
arrivals, gap onsets, gap closures, and completions, into a CSR event
FIFO, and the controller returns a victim index and logical DMA
descriptors. Physical address mapping, KV arrays, the HBM PHY, and
the on-chip network remain external. The IP therefore
owns~\eqref{eq:scoreq} and~\eqref{eq:budget}, not the bytes that
those decisions move.

Near-memory placement beside the hierarchy is motivated by the event
interface. A software scheduler that polls block-cache timestamps
approximates gap onset with millisecond granularity at best, and
Section~\ref{sec:sw-vllm} shows that this approximation inverts the
gap signal when tool waits approach the decode window. A controller
wired to the memory hierarchy observes the same events at cycle
granularity without software intervention, which is the structural
advantage that Section~\ref{sec:hw-place} quantifies under placement
variation.

\subsection{Unified Control Plane}
\label{sec:org}

Agent workloads interleave request and gap events at fine granularity
rather than in long homogeneous phases. A dual design that pairs a
\spear engine with a \tide engine under periodic synchronization is
the textbook decomposition, but it is structurally unsound under this
interleaving pattern. Eviction and promotion are two reads of one
ranking, not two independent computations, so a second engine is
either stale or redundant.

Let $R_{\mathrm{u}}$ denote the unified register file and
$R_{\mathrm{d}}$ the pair of private views in a dual design. The dual
stores
\begin{equation}
  \lvert R_{\mathrm{d}}\rvert=2\lvert R_{\mathrm{u}}\rvert
  \label{eq:rf}
\end{equation}
and applies a stale snapshot at every decision between synchronization
points. If the last sync was at event $e_{k-m}$, the dual victim is
\begin{equation}
  v_{\mathrm{d}}
  =\arg\max_{s\in\mathcal{L}}
    \mathrm{score}_{q}\!\bigl(s;\,R_{\mathrm{d}}(e_{k-m})\bigr),
  \label{eq:stale}
\end{equation}
whereas the unified victim uses $R_{\mathrm{u}}(e_{k})$.
Tab.~\ref{tab:org} confirms the structural prediction. Even with
every-event synchronization the dual design diverges on SRAM
placement, and with reduced synchronization the stale view causes
hit-rate drops of up to 5.4\,pp and inconsistent AMAT signs across
traces, confirming that a unified register file is not merely
preferable but structurally necessary.

\input{tables/tab5_org}

The unified pipeline therefore keeps one live register file and
sequences \esp before \okt in the same datapath, paying
$\lvert R_{\mathrm{u}}\rvert$ rather than~\eqref{eq:rf}.

\subsection{Fixed-Point Quantization of the Scoring Contract}
\label{sec:quant}

Realizing the floating-point policy in synthesizable logic requires an
integer image of~\eqref{eq:esp} whose ranking fidelity is
indistinguishable from the reference. The local complement
$\sigma(t)$ rather than the cumulative product-limit survival is the
key enabler, because a single ROM lookup avoids the sequential
dependence chain that a cumulative estimator would impose on the
pipeline. Let $b$ be a fractional width and $\Lambda_b=2^{b}-1$.
The unsigned quantizer
\begin{equation}
  Q_{b}(x)
  =\mathrm{clip}\!\left(
      \bigl\lfloor x\,\Lambda_b+\tfrac12\bigr\rfloor,\;
      0,\;\Lambda_b
    \right)
  \label{eq:uq}
\end{equation}
maps $x\in[0,1]$ onto $\mathrm{UQ}0.b$. Hazard and survival then
become
\begin{align}
  h_{q}
  &=Q_{b_{h}}\!\bigl(h(\min(t,50))\bigr),
  \label{eq:hq}
  \\
  \sigma_{q}
  &=\max\!\bigl(
      \bigl\lfloor\varepsilon\Lambda_{b_{h}}+\tfrac12\bigr\rfloor,\;
      \Lambda_{b_{h}}-h_{q}
    \bigr).
  \label{eq:survq}
\end{align}
A time divisor $\tau\in\{1,10^{3},10^{6}\}$ stores the gap as
\begin{equation}
  g_{q}
  =\mathrm{clip}\!\left(
      \bigl\lfloor g_s/\tau\bigr\rfloor,\;
      0,\;2^{b_{g}}-1
    \right),
  \label{eq:gapq}
\end{equation}
so $\tau=1$ is nanoseconds, $\tau=10^{3}$ is microseconds, and
$\tau=10^{6}$ is milliseconds. The two score terms are integers,
\begin{align}
  T_{1}
  &=\left\lfloor
      \frac{g_{q}\,\Lambda_{b_{h}}}{\sigma_{q}}
    \right\rfloor,
  \label{eq:t1}
  \\
  T_{2}
  &=\left\lfloor
      \frac{(\Lambda_{b_{h}}-\sigma_{q})\,P_{\mathbb{Z}}}
           {\Lambda_{b_{h}}}
    \right\rfloor,
  \label{eq:t2}
\end{align}
and the hardware score is the saturating sum
\begin{equation}
  \mathrm{score}_{q}(s)
  =\mathrm{sat}_{b_{s}}\!\bigl(T_{1}+T_{2}\bigr),
  \qquad
  b_{s}=64,
  \label{eq:scoreq}
\end{equation}
with $P_{\mathbb{Z}}=P/\tau$ so that the multiplicative penalty
remains an affine image of $(1-\sigma)P$ after the time scaling. The
EMA in~\eqref{eq:ema} is the same fraction in integer arithmetic,
\begin{equation}
  g_{q}
  \leftarrow
  \left\lfloor
    \frac{3\,g_{q}^{\mathrm{new}}+7\,g_{q}}{10}
  \right\rfloor.
  \label{eq:ema-int}
\end{equation}
A completed session writes $2^{b_{s}}-1$.

The contract is selected on a pre-registered grid of 180 format
points that vary $b_{h}\in\{8,10,12\}$,
$b_{g}\in\{24,32,64\}$, $\tau\in\{1,10^{3},10^{6}\}$, divider type,
and penalty form. Ranking fidelity is measured on 7265 real eviction
snapshots from the six traces and three capacity envelopes of
Section~\ref{sec:sweval}. The hard gates are top-1 victim agreement
${\ge}0.99$ and maximum hit-rate drift ${\le}0.20$\,pp. The sweep
selects
\begin{equation}
  (b_{h},b_{g},b_{s},\tau,P_{\mathbb{Z}})
  =(12,32,64,10^{3},5\times10^{8}),
  \label{eq:lock}
\end{equation}
at which the aggregate Kendall $\tau$ exceeds 0.998 and the worst-cell
hit-rate drift is 0.192\,pp. Twelve-bit hazard is the minimum width that passes the ranking gate,
microsecond storage balances register width against fidelity, and an
exact divider is preferred over a reciprocal LUT because the LUT ties
fidelity while adding a table of depth $\Lambda_{b_{h}}$. Section~\ref{sec:dse} shows that
the bits ranking demanded are essentially free relative to the session
register file.

\subsection{Event Core Microarchitecture}
\label{sec:micro}

The unified pipeline must evaluate~\eqref{eq:scoreq} for every live
session within the tool-gap window so that the eviction or migration
decision is ready before the next request arrives. A parallel
comparator tree would close in $O(1)$ cycles but scale as $O(N)$ in
area. A pipelined linear scan trades latency for area, closing in
$N{+}5$ cycles at a cost that grows with the register file rather
than with dedicated comparator logic. The five-stage depth maps
directly to the arithmetic of the quantized score, with the
critical-path divider of~\eqref{eq:t1} dictating the pipeline depth.

The stages are
\begin{equation}
  \begin{aligned}
    \mathrm{S0}&:\;
      (t,g_{q},\mathrm{done})\leftarrow R[s],\\
    \mathrm{S1}&:\;
      h_{q}\leftarrow\mathrm{ROM}[t],\;
      \sigma_{q}\leftarrow~\eqref{eq:survq},\\
    \mathrm{S2}&:\;
      T_{1}\leftarrow~\eqref{eq:t1},\\
    \mathrm{S3}&:\;
      T_{2}\leftarrow~\eqref{eq:t2},\\
    \mathrm{S4}&:\;
      \mathrm{score}_{q}\leftarrow~\eqref{eq:scoreq},\;
      \text{update running }\arg\max.
  \end{aligned}
  \label{eq:pipe}
\end{equation}
A full scan of $N$ live sessions takes
\begin{equation}
  L_{\mathrm{scan}}=N+5
  \label{eq:scan}
\end{equation}
cycles after pipeline fill, yielding a decision latency of
\begin{equation}
  \tau(N)=\frac{N+5}{f_{\mathrm{clk}}}.
  \label{eq:tau}
\end{equation}
The \tide budget uses the same microsecond gap,
\begin{equation}
  \mathcal{B}_{q}
  =\bigl\lfloor\Delta/\tau\bigr\rfloor\cdot B_{\mu},
  \label{eq:budgetq}
\end{equation}
where $B_{\mu}$ is tokens per microsecond.
Cosimulation of the locked format against the floating-point reference
is bit-exact on all 1\,937 events and 845 decisions of the GAIA/Gemma4
Edge-Tight trace.

Given the pipeline and the quantization contract, the remaining
architectural degree of freedom is the session capacity~$N$.

\subsection{Session-Capacity Design-Space Exploration}
\label{sec:dse}

Session capacity~$N$ governs both decision latency
through~\eqref{eq:tau} and register-file area. To evaluate the policy
side of this tradeoff, we define Agent Serving Effectiveness (ASE) as
the six-dataset geometric mean
\begin{equation}
  \mathrm{ASE}(N)=
  \Biggl(\prod_{d=1}^{6}
    \frac{\mathrm{HR}_d}{\mathrm{HR}_d^{\max}}
    \cdot
    \frac{\mathrm{AMAT}_d^{\min}}{\mathrm{AMAT}_d}
    \cdot
    \frac{\mathrm{TTFT}_d^{\min}}{\mathrm{TTFT}_d}
  \Biggr)^{1/18}
  \label{eq:ase}
\end{equation}
inside the capacity sweep, where the inner product runs over three
normalized metrics and the outer over six traces.

\begin{figure}[t]
\centering
\begin{subfigure}{0.49\linewidth}
\includegraphics[width=\linewidth]{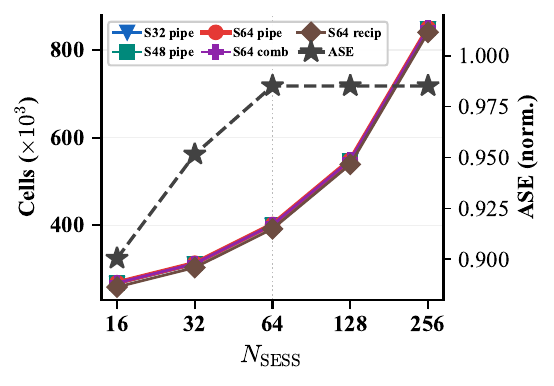}
\caption{Cell count and ASE versus $N_{\mathrm{sess}}$}
\label{fig:dse-2d}
\end{subfigure}\hfill
\begin{subfigure}{0.49\linewidth}
\includegraphics[width=\linewidth]{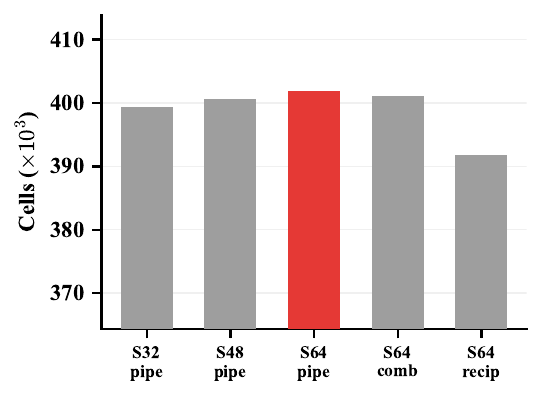}
\caption{Structural variants at $N{=}64$}
\label{fig:dse-n64}
\end{subfigure}
\caption{Design-space exploration of session capacity and structural
variants on the Yosys/Nangate45 synthesis flow.}
\label{fig:dse}
\end{figure}

Fig.~\ref{fig:dse} reports the exploration on the Yosys/Nangate45
flow used to rank variants by relative cell count. ASE saturates near
$N{=}64$ where trace concurrency plateaus, while decision latency
from~\eqref{eq:tau} remains well below the median tool gap at that
point. The capacity knee is therefore $N{=}64$. Absolute ASE values,
latency measurements, and the Design Compiler PPA at this point are
reported in Section~\ref{sec:hweval}.

Fig.~\ref{fig:dse-n64} details the $N{=}64$ column. The five live
configurations occupy a narrow cell-count band, and the selected
design point retains the exact divider of~\eqref{eq:t1} for bit-exact
scoring against the software reference, because the reciprocal-LUT
alternative ties ranking fidelity without improving the end-to-end
gate.

Sections~\ref{sec:sweval} and~\ref{sec:hweval} evaluate the policy
and report synthesis results.

%% file: tables/tab5_org.tex
\begin{table}[t]
\centering
\renewcommand{\arraystretch}{1.12}
\setlength{\tabcolsep}{3pt}
\scriptsize
\begin{threeparttable}
\caption{Architectural ablation of a unified control plane versus independent dual~IPs under varying synchronization granularity.}
\label{tab:org}
\begin{tabular*}{\columnwidth}{@{\extracolsep{\fill}}lrrrrrr@{}}
\toprule
\textbf{Trace} & \textbf{Vict$\neq$}\tnote{a} & \textbf{Prom$\neq$} &
\textbf{$\Delta$HR}\tnote{b} & \textbf{$\Delta$SRAM} &
\textbf{$\Delta$Pre} & \textbf{$\Delta$AMAT}\tnote{c} \\
 & (\%) & (\%) & (\%) & (\%) & (\%) & (\%) \\
\midrule
\rowcolor{gray!10}
\multicolumn{7}{c}{\textbf{Dual IPs copy $R$ after every request and gap}} \\
SWE/Qw3 & 0 & 0 & $+0.0$ & $+0.0$ & $+0.0$ & $+0.0$ \\
SWE/Dev & 0 & 0 & $+0.0$ & $+0.0$ & $+0.0$ & $+0.0$ \\
SWE/Ge4 & 0 & 0 & $+0.1$ & $+0.0$ & $-0.0$ & $+0.0$ \\
GAIA/Qw3 & 0 & 0 & $+0.0$ & $+0.5$ & $+0.8$ & $-1.7$ \\
GAIA/Dev & 0 & 0 & $+0.0$ & $-0.2$ & $+0.8$ & $-4.7$ \\
GAIA/Ge4 & 0 & 0 & $+0.2$ & $-1.3$ & $+1.0$ & $-1.8$ \\
\rowcolor{gray!10}
\multicolumn{7}{c}{\textbf{Dual IPs copy $R$ every 8 requests}} \\
SWE/Qw3 & 64 & 55 & $-2.2$ & $-0.4$ & $-1.2$ & $+1.2$ \\
SWE/Dev & 68 & 62 & $-1.9$ & $+0.0$ & $-1.1$ & $+1.1$ \\
SWE/Ge4 & 53 & 43 & $-5.4$ & $+0.0$ & $-1.9$ & $+2.0$ \\
GAIA/Qw3 & 35 & 50 & $-3.5$ & $-0.7$ & $-3.2$ & $+6.4$ \\
GAIA/Dev & 18 & 68 & $-0.2$ & $+1.0$ & $+0.9$ & $-5.0$ \\
GAIA/Ge4 & 39 & 45 & $-4.5$ & $-1.2$ & $-4.1$ & $+7.3$ \\
\rowcolor{gray!10}
\multicolumn{7}{c}{\textbf{Dual IPs copy $R$ only at gap start}} \\
SWE/Qw3 & 3 & 29 & $-0.0$ & $-0.1$ & $-0.0$ & $+0.0$ \\
SWE/Dev & 2 & 43 & $+0.2$ & $+0.2$ & $+0.0$ & $-0.0$ \\
SWE/Ge4 & 4 & 16 & $+0.4$ & $+0.0$ & $+0.1$ & $-0.1$ \\
GAIA/Qw3 & 5 & 39 & $+0.9$ & $-0.2$ & $+1.2$ & $-2.5$ \\
GAIA/Dev & 13 & 77 & $-3.4$ & $+0.7$ & $-1.7$ & $+9.0$ \\
GAIA/Ge4 & 4 & 29 & $+0.8$ & $-1.9$ & $+1.2$ & $-2.1$ \\
\bottomrule
\end{tabular*}
\begin{tablenotes}\scriptsize
\item[a] Vict$\neq$ and Prom$\neq$ are the shares of copied evictions and promotions that pick a different session.
\item[b] $\Delta$HR, $\Delta$SRAM, and $\Delta$Pre are percentage-point shifts versus the live table.
\item[c] $\Delta$AMAT is the percent AMAT change, and either sign represents a consistency failure.
\end{tablenotes}
\end{threeparttable}
\end{table}

%% file: sec_sweval.tex
\section{Software Evaluation}
\label{sec:sweval}

Before committing silicon area to the scheduling core of
Section~\ref{sec:arch}, the eviction and migration policies must be
validated on realistic agent traces. This section evaluates the
\unison policy along five dimensions that, taken together, stress
different failure modes of session-level caching.
Section~\ref{sec:sw-main} compares aggregate performance against
baseline and published policies on six traces.
Section~\ref{sec:sw-comp} ablates scoring and tiering to confirm their
complementarity. Section~\ref{sec:sw-signal} quantifies each signal
term via leave-one-out analysis on the full evaluation grid.
Section~\ref{sec:sw-robust} verifies that the gain persists under
storage-hierarchy perturbation, and Section~\ref{sec:sw-lut} examines
hazard lookup table (LUT) fairness and online convergence. Finally,
Section~\ref{sec:sw-vllm} replaces the trace-driven simulator with a
live vLLM server to validate that the policy transfers to a production
block-cache pool.

\subsection{Experimental Setup and Metrics}
\label{sec:sw-setup}

The evaluation replays six agent traces constructed by crossing the
SWE-bench~\cite{jimenez_swebench_2024} and
GAIA~\cite{mialon_gaia_2023} task suites with three model families,
namely Qwen3-Coder-30B~\cite{qwen3_2025},
Devstral-24B~\cite{devstral_2025}, and
Gemma4-E4B~\cite{gemma4_2026}, yielding 1\,415 sessions and 33\,596
turns in total. No public multi-agent serving trace with session-level
annotations exists as of this writing, so this single-agent replay
corpus is the most detailed open benchmark available. The main comparison includes recency (LRU), timeout,
the AGSERVE expected-completion estimator, CacheScout, \unison, and the
offline B\'el\'ady oracle. CacheScout is evaluated on a matched
two-tier harness at the Edge-Tight envelope against its own LRU
reference. Scoring-only and tiering-only variants are deferred to
Fig.~\ref{fig:ablation}. Six SRAM-and-HBM capacity envelopes span edge
through CXL configurations. Hit rate and prefill reduction are
arithmetic means over these envelopes, while AMAT/LRU and TTFT/LRU are
geometric-mean ratios relative to recency. TTFT is reported at the p50
percentile under $4\times$ load.

A request is an SRAM hit, an HBM hit, or a miss. With $R$ requests the
hit rate and the average memory access time are
\begin{equation}
  \mathrm{HR}=\frac{H_{\mathrm{S}}+H_{\mathrm{H}}}{R},
  \qquad
  \mathrm{AMAT}=\frac{1}{R}\sum_{i=1}^{R}
    \begin{cases}
      \ell_i\,t_{\mathrm{s}} & \text{SRAM hit},\\
      \ell_i\,t_{\mathrm{h}} & \text{HBM hit},\\
      \ell_i\,t_{\mathrm{p}} & \text{miss},
    \end{cases}
  \label{eq:hr-amat}
\end{equation}
where $t_{\mathrm{s}}$, $t_{\mathrm{h}}$, and $t_{\mathrm{p}}$ are the
per-token SRAM, HBM, and prefill costs. Prefill reduction is the
fraction of those tokens that a hit avoids,
\begin{equation}
  \mathrm{PreRED}=1-\frac{\sum_i \ell_i^{\mathrm{comp}}}{\sum_i \ell_i},
  \label{eq:prered}
\end{equation}
and the B\'el\'ady ratio is
\begin{equation}
  \mathrm{BR}
  =\frac{\mathrm{HR}}{\mathrm{HR}^{\star}}
  \label{eq:br}
\end{equation}
against the offline oracle. Prefetch accuracy is defined only when \tide
is present. Serving TTFT follows the same prefill accounting under a
trace-driven load model.

\subsection{Main Results}
\label{sec:sw-main}

Tab.~\ref{tab:main} compares \unison with baseline recency and timeout
rules, the AGSERVE expected-completion estimator, CacheScout, and the
offline B\'el\'ady oracle across all six traces. \unison is the best
non-oracle entry on hit rate and AMAT on every dataset, achieving a
B\'el\'ady ratio of $\mathrm{BR}{=}0.93$. It surpasses LRU on AMAT in
all 36 matched comparisons with a mean reduction of 34.8\% and
outperforms CacheScout on hit rate across all six datasets. The
remaining 7\% gap to B\'el\'ady reflects the inherent cost of
causality, since an online policy cannot observe the next arrival and
the residual is the portion that~\eqref{eq:esp} does not close.

\paragraph{TTFT and the hit-rate versus latency trade-off}
Although \unison dominates on hit rate and AMAT, the AGSERVE
expected-completion estimator achieves a marginally lower aggregate
TTFT/LRU ratio. The serving model charges leftover prefill tokens to
the request at the head of the GPU queue. The estimator evicts the
session it predicts will return latest and can therefore discard a
long-prefix session that the \unison completion penalty retains. The
retained prefixes improve hit rate and AMAT but leave a longer residual
prefill on the critical path of a competing arrival, allowing the
estimator to gain on first-token latency while losing on every other
metric. The
0.01-point TTFT difference therefore reflects a fundamental
hit-rate versus latency trade-off rather than a ranking deficiency.

More broadly, when long prefills saturate GPU compute as on
SWE/Qwen3 and SWE/Devstral, the bottleneck shifts to queuing delay and
policy ordering ceases to influence TTFT. The cache-bound regime where
eviction policy most directly improves latency is isolated in the vLLM
validation of Section~\ref{sec:sw-vllm}. Scoring and tiering ablations
are presented in Fig.~\ref{fig:ablation}.

\input{tables/tab4_main}

\subsection{Scoring and Tiering Ablation}
\label{sec:sw-comp}

\begin{figure*}[t]
\centering
\begin{subfigure}{0.24\textwidth}
\includegraphics[width=\linewidth]{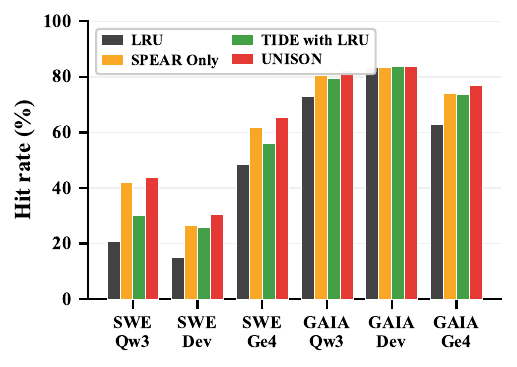}
\caption{Hit rate}
\label{fig:ablation-hr}
\end{subfigure}\hfill
\begin{subfigure}{0.24\textwidth}
\includegraphics[width=\linewidth]{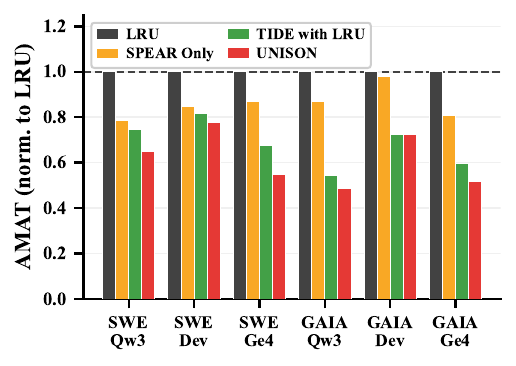}
\caption{AMAT / LRU}
\label{fig:ablation-amat}
\end{subfigure}\hfill
\begin{subfigure}{0.24\textwidth}
\includegraphics[width=\linewidth]{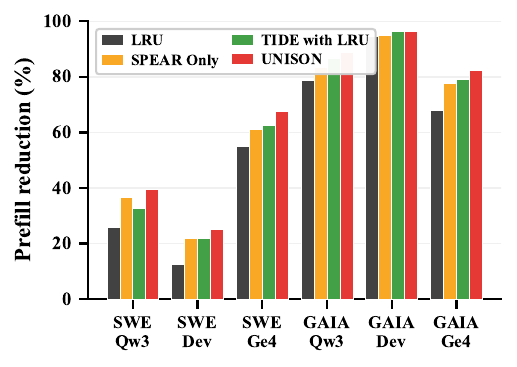}
\caption{Prefill reduction}
\label{fig:ablation-prered}
\end{subfigure}\hfill
\begin{subfigure}{0.24\textwidth}
\includegraphics[width=\linewidth]{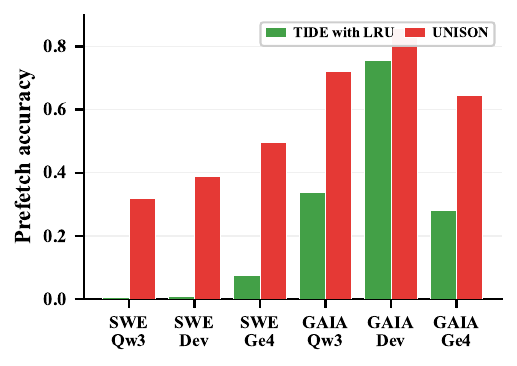}
\caption{Prefetch accuracy}
\label{fig:ablation-pfacc}
\end{subfigure}
\caption{Component ablation of scoring and tiering.}
\label{fig:ablation}
\end{figure*}

Scoring and tiering could in principle be substitutes, so
Fig.~\ref{fig:ablation} crosses both evictors with and without \tide to
test this hypothesis. The joint \unison configuration outperforms either
component in isolation because scoring determines which sessions to
retain while tiering determines where they reside. A strong evictor
without migration strands hot sessions in HBM, and a migrator without
\esp promotes the wrong sessions, yielding a prefetch accuracy of only
1.1\%.

\subsection{Signal Contribution Analysis}
\label{sec:sw-signal}

\begin{figure}[t]
\centering
\begin{subfigure}{0.48\linewidth}
\includegraphics[width=\linewidth]{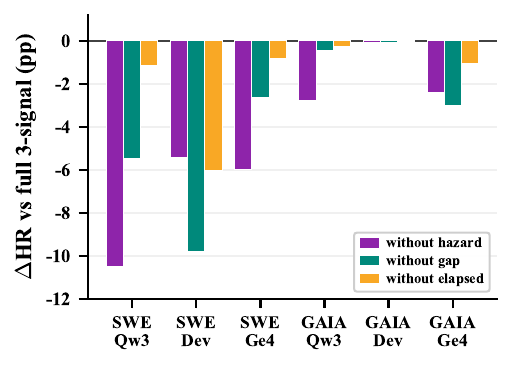}
\caption{$\Delta$HR vs full-3sig}
\label{fig:signal-hr}
\end{subfigure}\hfill
\begin{subfigure}{0.48\linewidth}
\includegraphics[width=\linewidth]{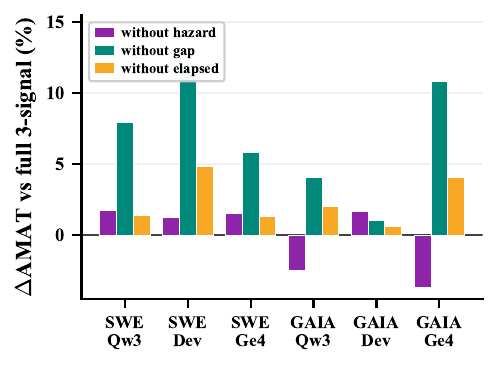}
\caption{$\Delta$AMAT vs full-3sig}
\label{fig:signal-amat}
\end{subfigure}
\caption{Mechanism-signal ablation of \spear via leave-one-out on the 36-point evaluation grid.}
\label{fig:signal}
\end{figure}

The hardware scorer of Section~\ref{sec:quant} retains only two of the
three \spear signals, so a leave-one-out analysis on the full 36-point
grid is needed to validate that reduction.
Fig.~\ref{fig:signal} confirms that hazard and gap recurrence are both
indispensable, while elapsed idle contributes the least with a mean
impact of 1.55\,pp on hit rate. Because elapsed idle is also the only
term that requires a per-slot timer in hardware, its omission from the
register-transfer level (RTL) implementation is justified on both
accuracy and area grounds.

\subsection{Robustness Across Capacity Operating Points}
\label{sec:sw-robust}

The preceding results use a single default storage envelope.
Tab.~\ref{tab:robust} verifies that the advantage is not an artifact of
that operating point by sweeping three hierarchy axes, namely HBM capacity
with SRAM fixed, SRAM capacity with HBM fixed, and the SRAM-to-HBM
partition with the total held at 60K tokens. The six-trace mean remains
positive on hit rate, AMAT, and prefill reduction at every tabulated
point. The gain grows monotonically with the SRAM share because a
larger fast tier amplifies the benefit of accurate residency ranking.
It diminishes only when HBM is large enough that recency alone
approaches the hit-rate ceiling, a regime in which two GAIA traces tie
on hit rate while the six-trace mean remains positive. Ancillary
parameters, including read latency, DMA bandwidth, and prefill rate, do
not reorder the policies and shift the mean AMAT gain by less than 0.6
percentage points around the 19.6\% baseline, so they are omitted from
the table.

\input{tables/tab6_robust}

\subsection{Hazard LUT Fairness and Online Convergence}
\label{sec:sw-lut}

All experiments so far rely on a hazard LUT fitted on the full trace,
which could bias the evaluation by overfitting its own sessions or by
embedding deployment-time knowledge unavailable during online serving.

Tab.~\ref{tab:lutcv} answers the first concern with a five-fold
cross-validation on SWE/Qwen3. Because a scatter of the same 20
fold-by-capacity cells would cluster on $y{=}x$, the table reports the
residual. The held-out LUT deviates from the all-session LUT by at
most 0.85\,pp and by 0.17\,pp on average, confirming that the hazard
distribution is a stable property of the turn-survival law rather than
of a particular session split.

\begin{figure}[t]
\centering
\includegraphics[width=\columnwidth]{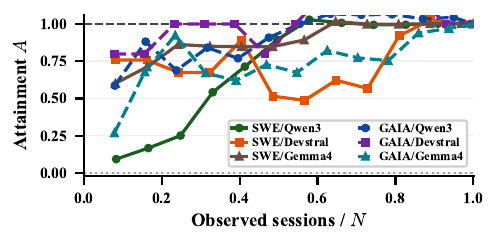}
\caption{Online convergence of the hazard mechanism.}
\label{fig:fair-online}
\end{figure}

Fig.~\ref{fig:fair-online} addresses the second concern by measuring
how quickly an initially empty LUT converges to the offline
performance. Each trace incrementally rebuilds the table as sessions
complete, accumulating the counts in~\eqref{eq:hazard} with $N$
ranging from 128 to 496. To compare traces with different absolute
hit-rate scales on a common frame, we define the attainment metric
\begin{equation}
  A(n)=\frac{H_{\mathrm{on}}(n)-H_{\mathrm{LRU}}}
            {H_{\mathrm{off}}-H_{\mathrm{LRU}}}
  \label{eq:attainment}
\end{equation}
as the fraction of the offline-minus-LRU hit-rate surplus realized
after $n$ observed sessions, where $A{=}0$ corresponds to the LRU
floor and $A{=}1$ to the offline table. All six convergence curves
reach $A{=}1$. Residual fluctuations on GAIA/Gemma4 reflect count
noise from its 128-session corpus, and the two Devstral traces start
near the ceiling because LRU already approaches the offline hit rate
on those workloads. The globally fitted LUT is therefore an evaluation
convenience rather than a hidden prior, as a production deployment
accumulates the same table incrementally at runtime.

\input{tables/tab_lut_cv}

\subsection{Production-Stack Validation with vLLM}
\label{sec:sw-vllm}

The preceding experiments use a trace-driven simulator that abstracts
away serving-stack scheduling and memory-management details. To
validate that the \spear ranking transfers to a real inference server,
we integrate the score into the vLLM~\cite{kwon_pagedattention_2023}
v1 prefix cache, replacing the default LRU victim order with the
\spear score derived exclusively from runtime-observable metadata.
Paired replays of the GAIA/Qwen3 trajectory issue token-identical loads
at 32-way concurrency and zero temperature. Tool-call frequency is
varied between tight and loose schedules to span the cache-contention
range that representative agent workloads exhibit. All pairings
complete without request errors, and Tab.~\ref{tab:vllm} reports the
measured gain against LRU under three deployment cases that each
isolate one operational factor.

\input{tables/tab7_vllm}

\paragraph{Gain on representative configurations}
The cache-bound case pairs a small testbed model whose fast decode
makes the workload prefill-dominated with a tight tool-call schedule;
\spear cuts mean TTFT by 35\% and tail end-to-end latency by 72\%. The
capacity-stressed case uses a loose schedule on a halved pool and
still delivers a 2.5\,pp hit-rate gain with a 6\% mean TTFT
reduction. The native-model case replaces the testbed with the
trajectory-native 30B under the tight schedule and confirms a positive
hit-rate gain of 1.6\,pp, ruling out a model-mismatch artifact.

\paragraph{Degradation under specific regimes}
Three traces not tabulated exhibit degraded or neutral outcomes.
On GAIA/Devstral under a halved pool and a tight schedule, tool gaps
fall below the decode latency, so the software idle timestamp inverts
the gap signal and hit rate drops by $6.6$\,pp. GAIA/Gemma4 has only
128 sessions, producing a noisy hazard LUT whose variance masks the
\spear advantage. SWE/Qwen3 is GPU-bound with negligible cache
pressure, matching the queue-bound SWE traces in
Tab.~\ref{tab:main}.

\paragraph{Efficacy boundary and hardware implication}
The degradation pattern reveals two efficacy boundaries of the
software patch. First, when the tool gap is shorter than the decode
window, a block-cache timestamp cannot resolve the idle onset and the
gap signal reverses, which is a fundamental limitation of any
software-level approximation that lacks a precise request-completion
event. Second, when prefill already saturates the GPU, a cache hit
saves tokens but does not shorten the queuing delay, so hit-rate gains
do not translate to latency gains. The first boundary is precisely what the event-driven hardware
interface of \unison eliminates, because a dedicated event FIFO stamps
gap onset and closure at cycle granularity and thereby removes the
software timing approximation that fails on short gaps. Note that the
\tide migration mechanism is not evaluated in this vLLM integration
because the production block cache exposes no tier-placement API, so
only the \spear ranking is exercised here.

%% file: tables/tab4_main.tex
\begin{table*}[t]
\centering
\renewcommand{\arraystretch}{1.15}
\setlength{\tabcolsep}{2pt}
\scriptsize
\begin{threeparttable}
\caption{Overall algorithmic performance of representative session-level policies across six agent-trace families.\tnote{a}}
\label{tab:main}
\begin{tabular*}{\textwidth}{@{\extracolsep{\fill}}lrrrrrrrrrrrrrrr@{}}
\toprule
\textbf{Policy} & \multicolumn{2}{c}{\textbf{SWE/Qw3}} & \multicolumn{2}{c}{\textbf{SWE/Dev}} & \multicolumn{2}{c}{\textbf{SWE/Ge4}} & \multicolumn{2}{c}{\textbf{GAIA/Qw3}} & \multicolumn{2}{c}{\textbf{GAIA/Dev}} & \multicolumn{2}{c}{\textbf{GAIA/Ge4}} & \multicolumn{3}{c}{\textbf{Aggregate}\tnote{b}} \\
\cmidrule(lr){2-3} \cmidrule(lr){4-5} \cmidrule(lr){6-7} \cmidrule(lr){8-9} \cmidrule(lr){10-11} \cmidrule(lr){12-13} \cmidrule(lr){14-16}
 & \textbf{HR} & \textbf{AMAT} & \textbf{HR} & \textbf{AMAT} & \textbf{HR} & \textbf{AMAT} & \textbf{HR} & \textbf{AMAT} & \textbf{HR} & \textbf{AMAT} & \textbf{HR} & \textbf{AMAT} & \textbf{BR} & \textbf{PreRED} & \textbf{TTFT}\tnote{c} \\
 & (\%) & /LRU & (\%) & /LRU & (\%) & /LRU & (\%) & /LRU & (\%) & /LRU & (\%) & /LRU & & (\%) & /LRU \\
\midrule
Recency (LRU) & 20.7 & 1.00 & 15.2 & 1.00 & 48.3 & 1.00 & 72.8 & 1.00 & 83.3 & 1.00 & 63.0 & 1.00 & 0.63 & 55.7 & 1.00 \\
Timeout (TTL) & 21.6 & 0.95 & 16.2 & 0.97 & 48.7 & 0.99 & 73.4 & 1.01 & 83.6 & 0.96 & 65.0 & 0.98 & 0.64 & 56.5 & 0.95 \\
ETA (AGSERVE)~\cite{agserve_2025} & 32.5 & 0.78 & 26.8 & 0.82 & 56.8 & 0.87 & 77.9 & 0.83 & 83.4 & 0.99 & 72.7 & 0.75 & 0.80 & 63.4 & \textbf{0.63} \\
CacheScout~\cite{cachescout_2026} & 2.6 & 1.00 & 2.8 & 1.00 & 4.4 & 1.00 & 31.9 & 0.95 & 67.2 & 0.98 & 24.0 & 0.98 & -- & 26.0 & 0.97 \\
\rowcolor{gray!10}
\textbf{\textsc{Unison}} & \textbf{43.8} & \textbf{0.65} & \textbf{30.5} & \textbf{0.78} & \textbf{65.3} & \textbf{0.55} & \textbf{82.3} & \textbf{0.49} & \textbf{83.6} & \textbf{0.72} & \textbf{76.7} & \textbf{0.52} & \textbf{0.93} & \textbf{66.6} & 0.64 \\
\textcolor{gray}{B\'el\'ady oracle} & 43.9 & 0.68 & 39.9 & 0.68 & 65.3 & 0.76 & 81.6 & 0.69 & 83.6 & 0.95 & 77.2 & 0.67 & 1.00 & 67.8 & 0.43 \\
\bottomrule
\end{tabular*}
\begin{tablenotes}\scriptsize
\item[a] Qw3 stands for Qwen3-Coder-30B, Dev stands for Devstral-24B and Ge4 stands for Gemma4-E4B.
\item[b] Hit rate(HR) and prefill reduction(PreRED) are means over six capacity envelopes.
\item[c] Average memory access time(AMAT/LRU) and time to first token(TTFT/LRU) metrics are geomean ratios to recency, and TTFT is p50 at $4\times$ load.
\end{tablenotes}
\end{threeparttable}
\end{table*}

%% file: tables/tab6_robust.tex
\begin{table}[t]
\centering
\renewcommand{\arraystretch}{1.12}
\setlength{\tabcolsep}{3pt}
\scriptsize
\begin{threeparttable}
\caption{Gain robustness of \unison against LRU under storage-hierarchy perturbation.}
\label{tab:robust}
\begin{tabular*}{\columnwidth}{@{\extracolsep{\fill}}lrrr@{}}
\toprule
 & \multicolumn{3}{c}{\textbf{Gain against LRU}\tnote{a}} \\
\cmidrule(lr){2-4}
\textbf{Point} & \textbf{$\Delta$HR}\tnote{b} & \textbf{$\Delta$AMAT} &
\textbf{$\Delta$Pre} \\
 & (\%) & (\%) & (\%) \\
\midrule
\rowcolor{gray!10}
\multicolumn{4}{c}{\textbf{HBM capacity, SRAM fixed at 10K tokens}} \\
30K & $+14.5$ & $+15.9$ & $+11.1$ \\
50K$^{\star}$ & $+15.5$ & $+19.6$ & $+12.4$ \\
100K & $+15.0$ & $+24.4$ & $+11.8$ \\
200K & $+10.4$ & $+16.2$ & $+5.3$ \\
\rowcolor{gray!10}
\multicolumn{4}{c}{\textbf{SRAM capacity, HBM fixed at 50K tokens}} \\
5K & $+13.0$ & $+15.9$ & $+8.7$ \\
10K$^{\star}$ & $+15.6$ & $+19.6$ & $+12.4$ \\
20K & $+18.4$ & $+26.5$ & $+15.2$ \\
40K & $+25.2$ & $+38.0$ & $+24.1$ \\
\rowcolor{gray!10}
\multicolumn{4}{c}{\textbf{SRAM:HBM partition, total fixed at 60K tokens}} \\
5K:55K & $+13.5$ & $+15.7$ & $+9.2$ \\
10K:50K$^{\star}$ & $+15.5$ & $+19.6$ & $+12.4$ \\
20K:40K & $+18.4$ & $+26.7$ & $+15.3$ \\
30K:30K & $+22.8$ & $+30.9$ & $+19.4$ \\
40K:20K & $+28.8$ & $+34.7$ & $+25.3$ \\
\bottomrule
\end{tabular*}
\begin{tablenotes}\scriptsize
\item[a] Each cell is the six-trace mean of the gain against LRU, with the starred row representing the Edge-Tight point.
\item[b] Hit-rate and prefill-reduction gains are percentage points, and the AMAT gain is the percent reduction in AMAT.
\end{tablenotes}
\end{threeparttable}
\end{table}

%% file: tables/tab_lut_cv.tex
\begin{table}[t]
\centering
\renewcommand{\arraystretch}{1.12}
\setlength{\tabcolsep}{4pt}
\scriptsize
\begin{threeparttable}
\caption{Cross-validation evidence for hazard-mechanism generalizability on SWE/Qwen3.}
\label{tab:lutcv}
\begin{tabular*}{\columnwidth}{@{\extracolsep{\fill}}lrrrr@{}}
\toprule
\textbf{Capacity} & \textbf{All-sess.\tnote{a}} & \textbf{Held-out\tnote{a}} &
\textbf{Mean $\Delta$ \tnote{b}} & \textbf{Max $\lvert\Delta\rvert$} \\
\midrule
100K & 4.90 & 4.88 & $-$0.03 & 0.11 \\
200K & 18.94 & 18.82 & $-$0.11 & 0.82 \\
300K & 38.75 & 38.53 & $-$0.21 & 0.78 \\
500K & 63.53 & 63.20 & $-$0.33 & 0.85 \\
\bottomrule
\end{tabular*}
\begin{tablenotes}\scriptsize
\item[a] The reported metric is five-fold mean hit rate in percent.
\item[b] $\Delta$ is held-out minus all-session in percentage points.
\end{tablenotes}
\end{threeparttable}
\end{table}

%% file: tables/tab7_vllm.tex
\begin{table}[t]
\centering
\renewcommand{\arraystretch}{1.12}
\setlength{\tabcolsep}{2pt}
\scriptsize
\begin{threeparttable}
\caption{Measured \spear gain under vLLM production-stack deployment.\tnote{a}}
\label{tab:vllm}
\begin{tabular*}{\columnwidth}{@{\extracolsep{\fill}}llllrrr@{}}
\toprule
\textbf{Case} & \textbf{Model} & \textbf{Pool} &
\textbf{Gap}\tnote{b} &
\textbf{$\Delta$HR} & \textbf{$\Delta$TTFT}\tnote{c} &
\textbf{$\Delta$e2e} \\
 & & & & (pp) & (\%) & (\%) \\
\midrule
Cache-bound & 1.5B & 60K & Tight & \textbf{+3.4} & $-35$ & $-72$ \\
Capacity-stressed & 1.5B & 30K & Loose & \textbf{+2.5} & $-6$ & $-4$ \\
Native-model & 30B & 60K & Tight & \textbf{+1.6} & $-1$ & $-3$ \\
\bottomrule
\end{tabular*}
\begin{tablenotes}\scriptsize
\item[a] Workload is GAIA/Qwen3 at 32-way concurrency.
1.5B is Qwen2.5-1.5B and 30B is the trajectory-native
Qwen3-Coder-30B-AWQ. Pool sizes are 60K and 30K tokens.
\item[b] Tight/Loose denotes the diverse tool-gap schedule.
\item[c] Deltas are versus paired LRU in vLLM and
TTFT is the mean and e2e is p99, in percent.
\end{tablenotes}
\end{threeparttable}
\end{table}

%% file: sec_hweval.tex
\section{Hardware Evaluation}
\label{sec:hweval}

Section~\ref{sec:arch} locked the integer contract, the unified
pipeline, and the session capacity at $N{=}64$. This section evaluates
the resulting IP under a 28-nm CMOS standard-cell flow, establishes the
architectural case for a dedicated scheduling core, and provides a
board-level demonstration. The evaluation proceeds from synthesis
results through efficiency, timeliness, and fidelity analyses to a
structural necessity argument that explains why the same policy cannot
be realized in software or as a pair of independent IPs.

\subsection{Locus Against Datapath KV Hardware}
\label{sec:hw-locus}

Table~\ref{tab:hw-kv} positions the \unison scheduling core against
token-level, device-level, and interconnect-level engines. Every
existing entry operates on a single request and optimizes a per-token
or per-block metric, whereas the scheduler ranks a pool of live
sessions on gap and completion signals that no token-level engine
observes. The two classes of hardware are complementary rather than
competing.

\input{tables/tab_hw_kv}

\subsection{Synthesis Methodology}
\label{sec:hw-method}

The synthesized netlist comprises an APB wrapper, a CSR file, an event
FIFO, and the unified \spear scan plus \tide engine, with the session
table mapped to flip-flops at roughly $N_{\mathrm{sess}}\times 126$
bits plus a four-bit \tide working copy. SRAM compiler macros, the HBM
PHY, and KV payload arrays are outside the cell area, and a prefix
directory together with a multi-instance router are likewise excluded.

Logic synthesis uses Synopsys Design Compiler on a commercial 28-nm
CMOS library at a 150\,MHz clock constraint, an application-side
budget rather than a race against the HBM PHY. Power is a vectorless
estimate at the timing corner. The RTL locks a pipelined scan and the
exact divider of~\eqref{eq:t1}, matching the bit-exact cosimulation of
Section~\ref{sec:micro}. Unless stated otherwise, bit-exact validation
and the FPGA demonstration use the GAIA/Gemma4 Edge-Tight trace, which
is the most hazard-sensitive of the six families and therefore the
hardest case for the fixed-point contract. Timeliness analysis draws on
all three GAIA families because only the GAIA traces carry recorded
tool-gap annotations.

\subsection{Area, Power, and Timing}
\label{sec:hw-ppa}

Table~\ref{tab:dc} reports the 64-session design point at the ASE knee
of Section~\ref{sec:dse}, occupying 0.169\,mm$^{2}$ at 13.6\,mW and
meeting 150\,MHz with 0.01\,ns positive setup slack for an estimated
$F_{\max}$ of 153.4\,MHz.

\input{tables/tab_dc}

Combinational logic accounts for 54\% of cell area and sequential
logic for 46\%. The \tide budget divider is the largest block at 34\%,
followed by the \spear divider at 17\% and the APB wrapper, CSR, and
event FIFO at a combined 6\%. The critical path runs through the first
\tide divider stage from the programmed DMA bandwidth register into the
remainder pipeline.

On the Yosys/Nangate45 flow of Fig.~\ref{fig:dse}, ASE reaches 0.900,
0.952, and 0.985 at 16, 32, and 64 sessions and saturates beyond 64
because trace concurrency never exceeds that count. Mean decision
latency per~\eqref{eq:tau} grows from 0.30\,$\mu$s at 16 sessions to
2.00\,$\mu$s at 64 and 3.95\,$\mu$s at 256. The five structural
variants at $N{=}64$ span 2.6\% in cells, and varying $b_{h}$ or
$b_{s}$ from the locked values shifts area by less than 3\%,
confirming that the precision demanded by ranking is essentially free
relative to the register file.

\subsection{Decision Efficiency and Control-Plane Overhead}
\label{sec:hw-eff}

The scheduler outputs a victim index and DMA descriptors rather than
computing attention, so its relevant efficiency metric is control-plane
cost relative to the data plane it manages.

\paragraph{Area amortization}
The 0.169\,mm$^{2}$ controller manages a two-tier pool of 10\,K to
60\,K tokens. At 128 heads, 128 dimensions, and 16-bit KV per layer, a
60\,K-token HBM bank occupies several hundred megabytes, and the
scheduling core is negligible relative to the HBM PHY and KV payload
arrays that together dominate die area on any contemporary inference
accelerator. Yet this vanishing fraction of total silicon delivers a
34.8\% mean AMAT reduction and a B\'el\'ady ratio of 0.93, a
cost-benefit asymmetry analogous to the branch predictor that occupies
less than 2\% of core area while eliminating the majority of pipeline
stalls.

\paragraph{Energy per decision}
At 13.6\,mW and a 2.00\,$\mu$s mean decision latency, each eviction
or migration decision consumes approximately 27.2\,pJ, which is orders
of magnitude below the per-access energy of the HBM transactions that
the decision orchestrates.

\paragraph{Workload proportionality}
Controller area scales with $N_{\mathrm{sess}}$ through the register
file, not with model dimensionality or sequence length. Increasing model
size from 7\,B to 70\,B parameters multiplies the KV store and
attention datapath but leaves the scheduling core unchanged, while
doubling session capacity beyond 64 roughly doubles the register file
without improving ASE, which is already saturated at 0.985. A
general-purpose microcontroller could in principle execute the same
algorithm, but its instruction-fetch and interrupt overhead would push
decision latency into the tens of microseconds, and the area of even a
minimal core with tightly coupled memory exceeds the 0.169\,mm$^{2}$
of the dedicated datapath while delivering lower determinism.

\subsection{Decision Timeliness}
\label{sec:hw-time}

At $N{=}64$ and 150\,MHz the mean scan latency is 2.00\,$\mu$s and
the worst case is 3.14\,$\mu$s. The GAIA traces exhibit median tool
gaps of 3.7\,s for Qwen3, 4.4\,s for Devstral, and 7.2\,s for
Gemma4, with the minimum observed gap across all three families at
1\,ms. The worst-case decision latency is therefore 0.31\% of the
minimum gap and below 0.0001\% of the median, leaving four to six
orders of magnitude of timing headroom.

This margin implies that even a ten-fold frequency reduction to
15\,MHz would keep the decision latency at 31\,$\mu$s, well within the
millisecond-scale minimum gap, so the controller can be placed in a
slow clock domain or power-gated between events. By contrast, a
software scheduler polling block-cache timestamps at millisecond
granularity occupies the same order of magnitude as the minimum gap
itself, which is why the software approximation inverts the gap signal
on short waits as documented in Section~\ref{sec:sw-vllm}.

\subsection{Implementation Fidelity}
\label{sec:hw-fid}

Cosimulation of the locked RTL format against the floating-point
reference is bit-exact on all 1\,937 events and 845 decisions of the
validation trace. Across 7\,265 real eviction snapshots from six traces and
three capacity envelopes, the aggregate top-1 victim agreement is
0.9945, the Kendall $\tau$ exceeds 0.998, and per-dataset agreement
ranges from 0.991 on SWE/Devstral at 2\,400 snapshots to 1.000 on
GAIA/Devstral at 16 snapshots, with no dataset falling below the 0.99
hard gate. The worst-cell end-to-end hit-rate drift is 0.192\,pp
within the 0.20\,pp hard gate, and the mean drift across all 18
dataset-by-capacity cells is 0.019\,pp while the AMAT drift is
$-0.018$\%.

Because each eviction decision is an independent $\arg\max$ over
instantaneous scores, the 0.55\% top-1 disagreement is a per-snapshot
coin flip that does not accumulate into trajectory-level drift, as
confirmed by the 0.019\,pp mean hit-rate deviation over full traces.
The fixed-point pipeline is therefore functionally transparent with
respect to the policy it implements, reproducing the ranking within a
residual that is smaller than the inter-policy gap between \unison and
CacheScout on every dataset.

\subsection{Structural Necessity of the Unified Near-Memory Design}
\label{sec:hw-nec}

Three lines of evidence converge to show that the same policy cannot be
realized more cheaply in software or as a pair of independent IPs.

\paragraph{Software cannot reconstruct the event interface}
Section~\ref{sec:sw-vllm} demonstrates that a software scheduler
operating on block-cache timestamps inverts the gap signal when the
tool wait approaches the decode window, because the timestamp
granularity at millisecond resolution is coarser than the gap itself,
whereas the hardware event FIFO resolves the same transition at cycle
granularity. More broadly, any software scheduler sharing the Python
runtime is subject to GIL serialization and OS-level jitter that
inflate scheduling latency by orders of magnitude beyond the
microsecond decisions the hardware achieves.

\paragraph{A dual-IP design diverges under agent interleaving}
Section~\ref{sec:org} shows that splitting eviction and migration into
two engines with private register files causes hit-rate drops of up to
5.4\,pp even under every-event synchronization, because agent workloads
interleave request and gap events at fine granularity rather than in
long homogeneous phases.

\paragraph{Convergence of the three constraints}
Cycle-level event observation requires near-memory placement, consistent
state requires a unified register file, and deterministic gap-window
completion requires a pipelined datapath. Relaxing any one constraint
re-introduces a documented failure mode, so the architectural
contribution is this constraint intersection realized as a
sub-0.2\,mm$^{2}$ control-plane IP.

\subsection{Board-Level Demonstration}
\label{sec:hw-fpga}

\begin{figure}[t]
\centering
\includegraphics[width=\columnwidth]{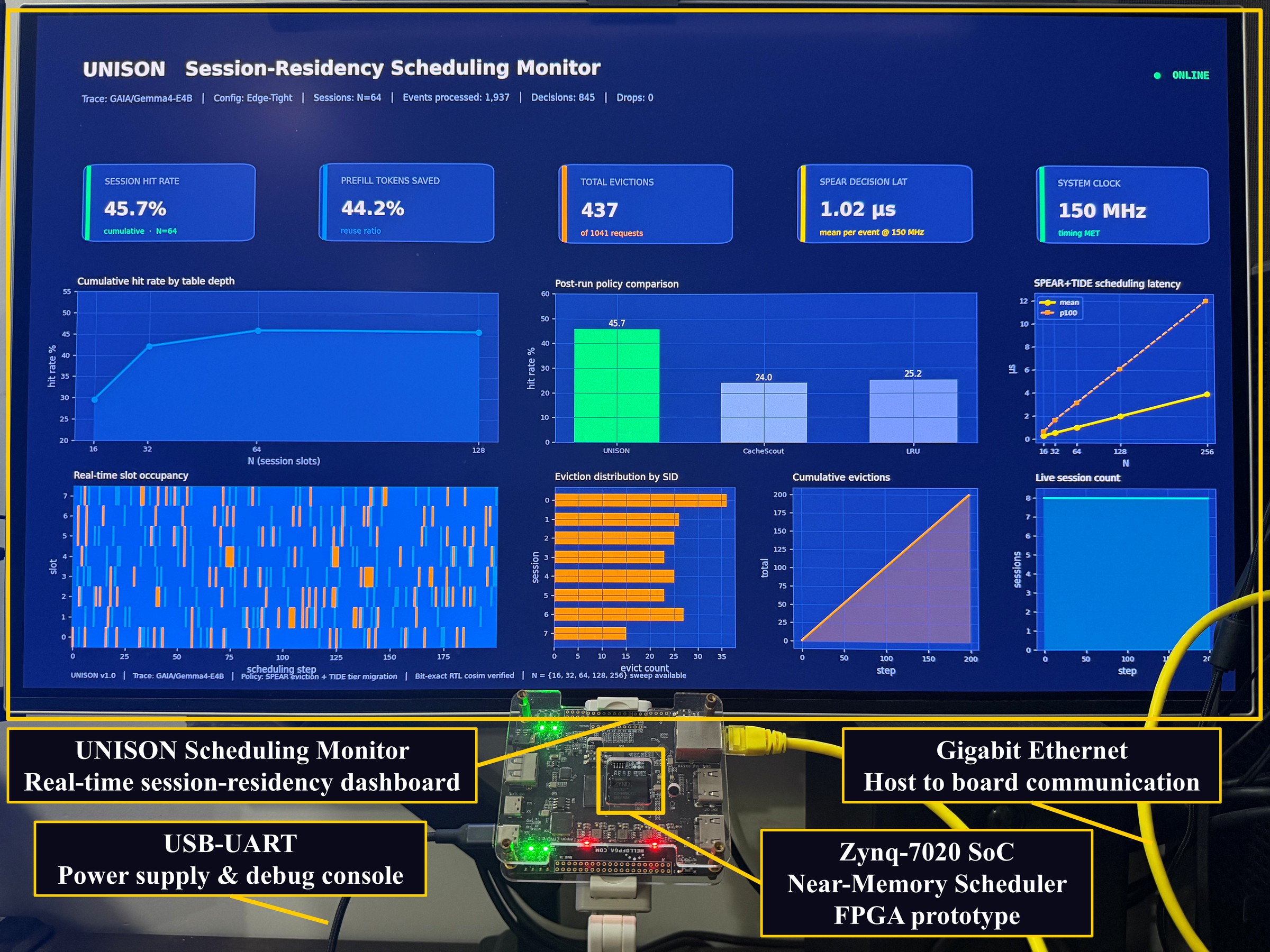}
\caption{FPGA prototype implementation of the \unison scheduling core
on Xilinx Zynq-7020.}
\label{fig:fpga}
\end{figure}

Fig.~\ref{fig:fpga} shows a Xilinx Zynq-7020 prototype that replays
the validation trace over 1\,937 events and 845 decisions, stamping
decision latency at the 150\,MHz Design Compiler clock. The
post-implementation netlist occupies 43\,415 LUTs at 81.6\% slice
utilization, 20\,938 flip-flops at 19.7\%, and 20 DSP48E1 blocks at
9.1\%, with no block RAM consumed.

\subsection{Physical Placement Considerations}
\label{sec:hw-place}

The timeliness margin of Section~\ref{sec:hw-time} implies that the
controller does not require co-location with the HBM PHY clock domain.
The essential requirement is access to the event stream that software
timestamps cannot reconstruct, which a sideband event bus from the
memory controller provides without a shared clock or a high-bandwidth
data interface.

%% file: tables/tab_hw_kv.tex
\begin{table*}[t]
\centering
\renewcommand{\arraystretch}{1.12}
\setlength{\tabcolsep}{2pt}
\scriptsize
\begin{threeparttable}
\caption{Comparison of representative KV cache hardware against the
session-level near-memory scheduler.\tnote{a}}
\label{tab:hw-kv}
\begin{tabular*}{\textwidth}{@{\extracolsep{\fill}}lllll@{}}
\toprule
\textbf{Work} & \textbf{Granularity} & \textbf{Locus} &
\textbf{Decision epoch} & \textbf{Mechanism} \\
\midrule
Token-Picker~\cite{park_tokenpicker_2024} & token & attention datapath
  & decode step & lossy probability prune; skip KV transfer \\
UniCAIM~\cite{xu_unicaim_2025} & token & CAM/CIM array
  & attention step & lossy static-dynamic prune; in-place attention \\
KV-MMU~\cite{moradifirouzabadi_kvmmu_2025} & token & accelerator MMU
  & generate step & lossy regularized token replace \\
HiKV~\cite{fang_hikv_2026} & token and element & attention datapath
  & decode step & lossy two-stage importance sort \\
Kelle~\cite{xia_kelle_2025} & token & eDRAM controller
  & decode step & lossy attention-score evict; refresh density \\
CXL-SpecKV~\cite{liu_cxlspeckv_2026} & block & CXL FPGA
  & speculative prefetch & expand pool; predict and move blocks \\
V-Rex~\cite{kim_vrex_2026} & video token & retrieval engine
  & frame / iterative prefill & retrieve a frame subset \\
\rowcolor{gray!10}
\textbf{\textsc{Unison}} & session & near-memory scheduler
  & request and gap events & lossless next-reference rank; evict and tier \\
\bottomrule
\end{tabular*}
\begin{tablenotes}\scriptsize
\item[a] Token-level importance datapaths also include
MATA~\cite{zhu_mata_2026} and VEDA~\cite{wang_veda_2025}.
\end{tablenotes}
\end{threeparttable}
\end{table*}

%% file: tables/tab_dc.tex
\begin{table}[t]
\centering
\renewcommand{\arraystretch}{1.12}
\setlength{\tabcolsep}{3pt}
\scriptsize
\begin{threeparttable}
\caption{Back-end synthesis results of the \unison scheduling core.}
\label{tab:dc}
\begin{tabular*}{\columnwidth}{@{\extracolsep{\fill}}lrrrrr@{}}
\toprule
\textbf{Process} & \textbf{Area}\tnote{a} &
\textbf{Power} & \textbf{WNS} & \textbf{$F_{\max}$} &
\textbf{Cells} \\
 & (mm$^{2}$) & (mW) & (ns) & (MHz) &  \\
\midrule
28-nm CMOS & 0.169 & 13.6 & +0.01 & 153.4 & 158k \\
\bottomrule
\end{tabular*}
\begin{tablenotes}\scriptsize
\item[a] Scope is the
APB wrapper, CSR, event FIFO, and unified core, excluding HBM PHY and KV payload SRAM.
\end{tablenotes}
\end{threeparttable}
\end{table}

%% file: sec_conc.tex
\section{Conclusion}
\label{sec:conc}

As large language models are composed into agents that retain a
growing key-value (KV) cache across tool waits, many concurrent
sessions share a finite static random-access memory (SRAM) and
high-bandwidth memory (HBM) pool, so that eviction and hierarchical
placement become a session-level efficiency problem orthogonal to
compute-mode optimization. Existing proxies based on recency,
timeout, or identity miss the mechanism information of the loop and
therefore treat a live wait as a cold, discardable unit.
\unison addresses both decisions from runtime-observable signals.
\spear ranks who leaves from a gap exponential moving average and a
turn-indexed hazard, while \tide spends the observed wait as a
direct memory access (DMA) budget for who sits in the fast tier, and
the two modules share one live ranking in an event-driven
near-memory scheduler beside the hierarchy.
Evaluated on six traces crossing
SWE-bench~\cite{jimenez_swebench_2024} and
GAIA~\cite{mialon_gaia_2023} with three model
families, totaling 1\,415 sessions, 33\,596 turns, and six capacity
envelopes from edge to cloud, the joint policy is the best
non-oracle entry on hit rate and average memory access time (AMAT)
on every trace, raising the hit rate by $0.3\%$ to $23.1\%$ and
reducing AMAT by $22\%$ to $51\%$. On the long-horizon traces it
also lowers serving time to first token (TTFT) by $58\%$ to $89\%$
when extra prefill sits on the critical path. The same ranking is realized as an event-driven near-memory
scheduler that occupies 0.169\,mm$^{2}$ and draws 13.6\,mW at
150\,MHz for 64 sessions in 28-nm CMOS.